\documentclass[aps,prd,twocolumn,superscriptaddress,floatfix,nofootinbib,natbib]{revtex4-1}
\usepackage{bm}
\usepackage{graphicx}
\usepackage[large]{subfigure}
\usepackage{amssymb, amsmath}
\usepackage[amssymb]{SIunits}
\usepackage{tabularx}
\usepackage{enumerate}
\usepackage{enumitem}
\usepackage{color}
\usepackage{footnote}
\usepackage[colorlinks,linkcolor=blue,citecolor=blue,urlcolor=blue ]{hyperref}

\begin{document}

\title{Cosmological Parameters from pre-Planck CMB Measurements: a 2017 Update}
\author{Erminia~Calabrese}\affiliation{Sub-department of Astrophysics, University of Oxford, Keble Road, Oxford OX1 3RH, UK}
\author{Ren\'ee~A.~Hlo\v{z}ek}\affiliation{Dunlap Institute and Department of Astronomy and Astrophysics, University of Toronto, 50 St George Street, Toronto M5S 3H4, Canada}
\author{J.~Richard~Bond}\affiliation{CITA, University of Toronto, Toronto, ON M5S 3H8, Canada}
\author{Mark~J.~Devlin}\affiliation{Department of Physics and Astronomy, University of Pennsylvania, 209 South 33rd St., Philadelphia,PA 19104,USA}
\author{Joanna~Dunkley}\affiliation{Joseph Henry Laboratories of Physics, Jadwin Hall, Princeton University, Princeton, NJ 08544,USA} \affiliation{Dept. of Astrophysical Sciences, Peyton Hall, Princeton University, Princeton, NJ 08544, USA}
\author{Mark~Halpern}\affiliation{Department of Physics and Astronomy, University of British Columbia, Vancouver, BC V6T 1Z4, Canada}
\author{Adam~D.~Hincks}\affiliation{Department of Physics, University of Rome ``La Sapienza'', Piazzale Aldo Moro 5, I-00185 Rome, Italy} 
\author{Kent~D.~Irwin}\affiliation{Department of Physics, Stanford University, Stanford, CA, 94305, USA}
\author{Arthur~Kosowsky}\affiliation{Department of Physics and Astronomy, University of Pittsburgh, Pittsburgh, PA 15260, USA}
\author{Kavilan~Moodley}\affiliation{Astrophysics and Cosmology Research Unit, School of Mathematics, Statistics and Computer Science, University of KwaZulu-Natal, Durban, 4041, South Africa}
\author{Laura~B.~Newburgh}\affiliation{Department of Physics, Yale University, New Haven, CT 06520, USA}
\author{Michael~D.~Niemack}\affiliation{Department of Physics, Cornell University, Ithaca, NY, USA 14853}
\author{Lyman~A.~Page}\affiliation{Joseph Henry Laboratories of Physics, Jadwin Hall, Princeton University, Princeton, NJ 08544,USA}  
\author{Blake~D.~Sherwin}\affiliation{Berkeley Center for Cosmological Physics, Lawrence Berkeley National Laboratory, Berkeley, CA, 94720, USA}  
\author{Jonathan~L.~Sievers}\affiliation{Astrophysics and Cosmology Research Unit, School of Chemistry and Physics, University of KwaZulu-Natal, Durban 4041, South Africa}\affiliation{National Institute for Theoretical Physics (NITheP), KZN
node, Durban 4001, South Africa}
\author{David~N.~Spergel}\affiliation{Dept. of Astrophysical Sciences, Peyton Hall, Princeton University, Princeton, NJ 08544, USA}\affiliation{Center for Computational Astrophysics, Flatiron Institute, New York, NY 10010, USA}
\author{Suzanne~T.~Staggs}\affiliation{Joseph Henry Laboratories of Physics, Jadwin Hall, Princeton University, Princeton, NJ 08544,USA}
\author{Edward~J.~Wollack}\affiliation{NASA/Goddard Space Flight Center, Greenbelt, MD 20771, USA}

\begin{abstract}
We present cosmological constraints from the combination of the full mission 9-year WMAP release and small-scale temperature data from the pre-\textit{Planck} ACT and SPT generation of instruments. This is an update of the analysis presented in Calabrese et al. 2013 and highlights the impact on $\Lambda$CDM cosmology of a 0.06~eV massive neutrino -- which was assumed in the \textit{Planck} analysis but not in the ACT/SPT analyses --  and a \textit{Planck}-cleaned measurement of the optical depth to reionization. We show that cosmological constraints are now strong enough that small differences in assumptions about reionization and neutrino mass give systematic differences which are clearly detectable in the data. We recommend that these updated results be used when comparing cosmological constraints from WMAP, ACT and SPT with other surveys or with current and future full-mission \textit{Planck} cosmology. Cosmological parameter chains are publicly available on the NASA's LAMBDA data archive.
\end{abstract}
\maketitle

\section{Introduction}

\begin{table*}[htb!]
\centering
\caption{Standard $\Lambda$CDM parameters with 68\% confidence level from the combination of WMAP9, ACT and SPT, and including a $\tau$ prior folding in recent \textit{Planck} HFI measurements~\cite{Plancktau2016}. The last two columns report a direct comparison with constraints from \textit{Planck} 2015 data derived with the same $\tau$ prior. \label{tbl1}}
\centering
\begin{tabular}{lcccccc}
\hline\hline
 Parameter & WMAP9 & WMAP9 & WMAP9 & (2017)-(2013) & PlanckTT & PlanckTTTEEE\\
 & +ACT & +SPT & +ACT+SPT & [in units of $\sigma$] & & \\
\hline
$100\Omega_b h^2$ & $2.243\pm0.040$ & $2.223\pm0.033$ & $2.242\pm0.032$ & $-0.25$ & $2.217\pm0.021$& $2.222\pm0.015$\\
$100\Omega_c h^2$ & $11.56\pm0.43$ &  $11.26\pm0.36$ & $11.34\pm0.36$ & $+0.41$ & $12.05\pm0.21$ & $12.03\pm0.14$\\
$10^4 \theta$ & $103.95\pm0.19$ &  $104.23\pm0.10$ & $104.24\pm0.10$ & $-0.21$ & $104.078\pm0.047$ & $104.069\pm0.032$\\
$\tau$ & $0.060\pm0.009$ & $0.057\pm0.009$ & $0.058\pm0.009$ & $-2.1$ & $0.064\pm0.010$ & $0.065\pm0.009$\\
$n_s$  & $0.966\pm0.010$ &  $0.9610\pm0.0089$ & $0.9638\pm0.0087$ & $-0.37$ & $0.9625\pm0.0056$ & $0.9626\pm0.0044$\\
ln$(10^{10} A_s)$ & $3.037\pm0.023$ & $3.018\pm0.021$ & $3.025\pm0.021$ & $-1.9$ & $3.064\pm0.020$ & $3.067\pm0.019$\\
\hline
$\Omega_\Lambda$$^a$  & $0.703\pm0.025$ & $0.726\pm0.019$ & $0.723\pm0.019$ & $-0.71$ & $0.680\pm0.013$ & $0.6812\pm0.0086$\\
$\Omega_{\rm m}$  & $0.296\pm0.025$ & $0.273\pm0.019$ & $0.277\pm0.019$ & $+0.71$ & $0.320\pm0.013$ & $0.3188\pm0.0086$\\
$\sigma_8$& $0.792\pm0.020$ & $0.774\pm0.018$ & $0.780\pm0.017$ & $-1.5$ & $0.820\pm0.010$ & $0.8212\pm0.0086$\\
$t_0$ & $13.813\pm0.093$ & $13.729\pm0.063$ & $13.715\pm0.062$ & $+0.81$ & $13.823\pm0.035$ & $13.822\pm0.025$\\
$H_0 $ & $68.5\pm2.0$ & $70.5\pm1.6$ & $70.3\pm1.6$ & $-0.74$ & $67.00\pm0.90$ & $67.03\pm0.61$\\
\hline\hline
\footnotetext[1]{Derived parameters: Dark energy density, total matter density, the amplitude of matter fluctuations on 8 $h^{-1} \rm{Mpc}$ scales, the age of the Universe in \rm{Gyr}, the Hubble constant in units of \rm{km/s/Mpc}.}
\end{tabular}
\end{table*}

In Calabrese et al. 2013 \cite{Calabrese2013} (hereafter C13) we presented cosmological results from the combination of Cosmic Microwave Background (CMB) experiments preceding the first data release of the \textit{Planck} satellite \cite{PlanckI2013}. We used data from the WMAP satellite after the completion of the mission including 9 years of large-scale CMB temperature and polarization observations \cite{Bennett2013,Hinshaw2013}, complemented by the final release of small-scale CMB temperature observations from the first generation of instruments of the Atacama Cosmology Telescope (ACT) \cite{Das2013,Dunkley2013,Sievers2013} and the South Pole Telescope (SPT) \cite{Story2013,Hou2014}. 

The results in C13 are sometimes used in comparison with the cosmological constraints obtained by the \textit{Planck} satellite, with low-redshift cosmology from galaxy surveys, and with local measurements of the expansion rate of the Universe (see e.g., \cite{Spergel2015,Anderson2014,Delubac2015,Addison2016,Riess2016,Mueller2016,Joudaki2016}).
In this brief note we want to highlight that cosmological constraints are now strong enough that small differences in assumptions about reionization and neutrino mass give systematic differences which are clearly detectable in the data.
We show that to have a direct comparison to \textit{Planck} cosmology \cite{PlanckXVI2013,PlanckXIII2015} or to galaxy constraints on the matter density and amplitude of matter fluctuations, two main things need to be updated in the C13 analysis:
\begin{enumerate}[leftmargin=0.5cm,noitemsep,nolistsep]
\item Starting in 2013, following the \textit{Planck} analyses~\cite{PlanckXVI2013}, estimates of cosmological parameters assume as baseline in a $\Lambda$CDM model a non-zero neutrino mass of 0.06~eV; this was not the case in C13 where neutrinos were treated as relativistic particles.
\item A reanalysis of the large-scale WMAP polarization by the \textit{Planck} team, using the new \textit{Planck} 353~GHz channel as thermal dust tracer, highlighted residual foreground contamination in the WMAP data leading to a $1\sigma$ bias in the estimate of the optical depth to reionization parameter, $\tau$ (see discussion in Refs.~\cite{PlanckXI2015,PlanckXIII2015}). A new and tighter measurement of $\tau$ has now been derived with \textit{Planck} HFI data~\cite{Plancktau2016} and should replace the WMAP one used in C13.\\
Although this \textit{Planck} result came after the WMAP/ACT/SPT measurements, it informs the comparison and so its effects are included here. We note that $\tau$ is the most uncertain of the cosmological parameters and the most likely to evolve with future measurements.\\
\end{enumerate}

In light of this, we present here a revised cosmology from WMAP9+ACT+SPT and recommend that these updated constraints be used for comparisons with other surveys. The cosmological parameter chains are available on the NASA's LAMBDA data archive at \url{http://lambda.gsfc.nasa.gov/product/act/act_prod_table.cfm}. The likelihood code for ACT and SPT is the same used in C13 and is available on LAMBDA (\url{http://lambda.gsfc.nasa.gov/product/act/}) and on the ACT website (\url{http://www.physics.princeton.edu/act/}).

\section{Analysis and Results}\label{sec:method} 
\begin{figure}[ht!]
\center
\includegraphics[width=\columnwidth]{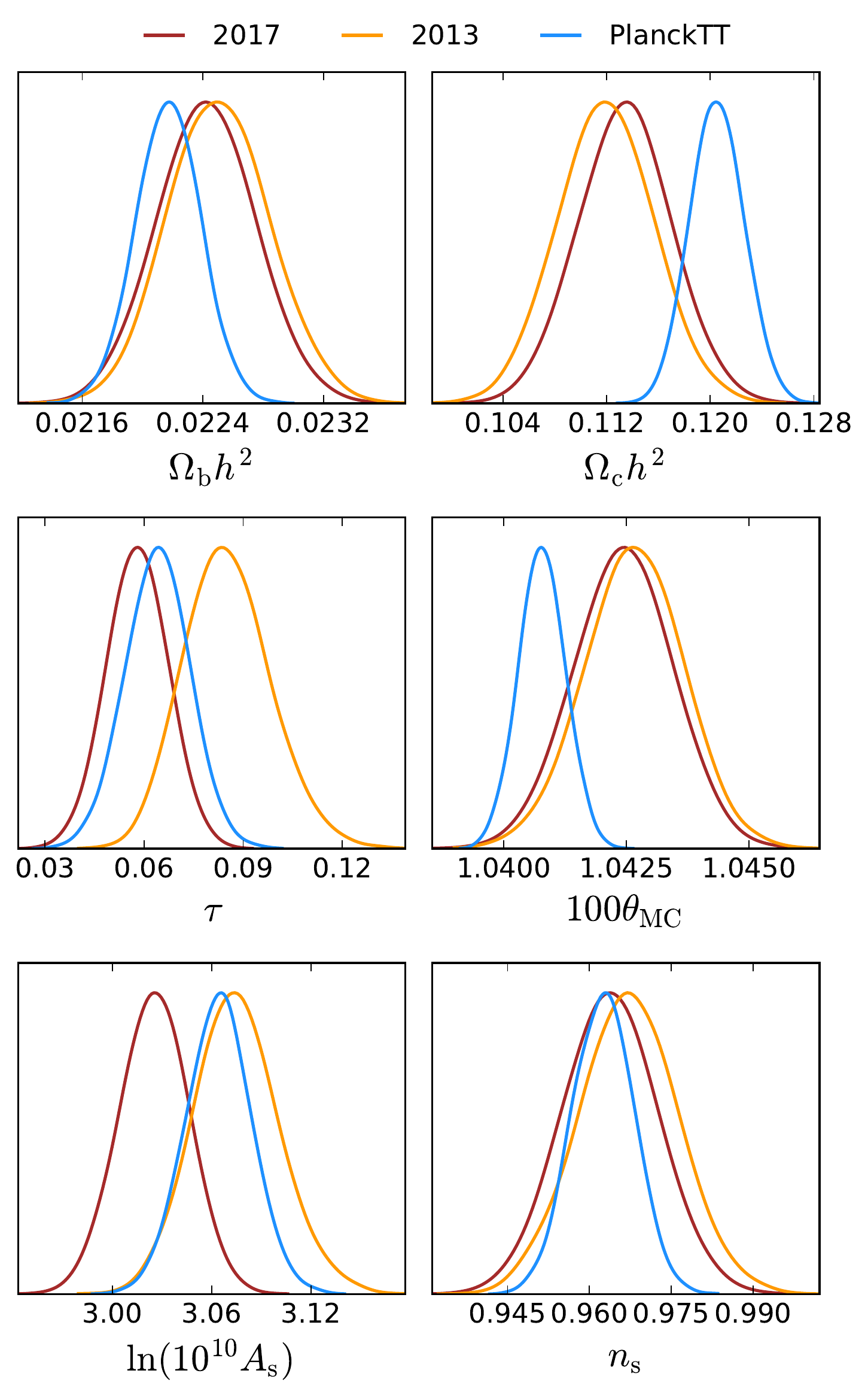}
\caption{One-dimensional posterior distributions for the six basic $\Lambda$CDM parameters as obtained in Calabrese et al. 2013 and with this revised analysis. We also show a direct comparison with the constraints from the \textit{Planck} baseline dataset~\cite{PlanckXIII2015}, analysed with the same assumptions (see also Table~\ref{tbl1}).
\label{fig:lcdm}
}
\end{figure}

\begin{figure}[ht!]
\center
\includegraphics[scale=0.6]{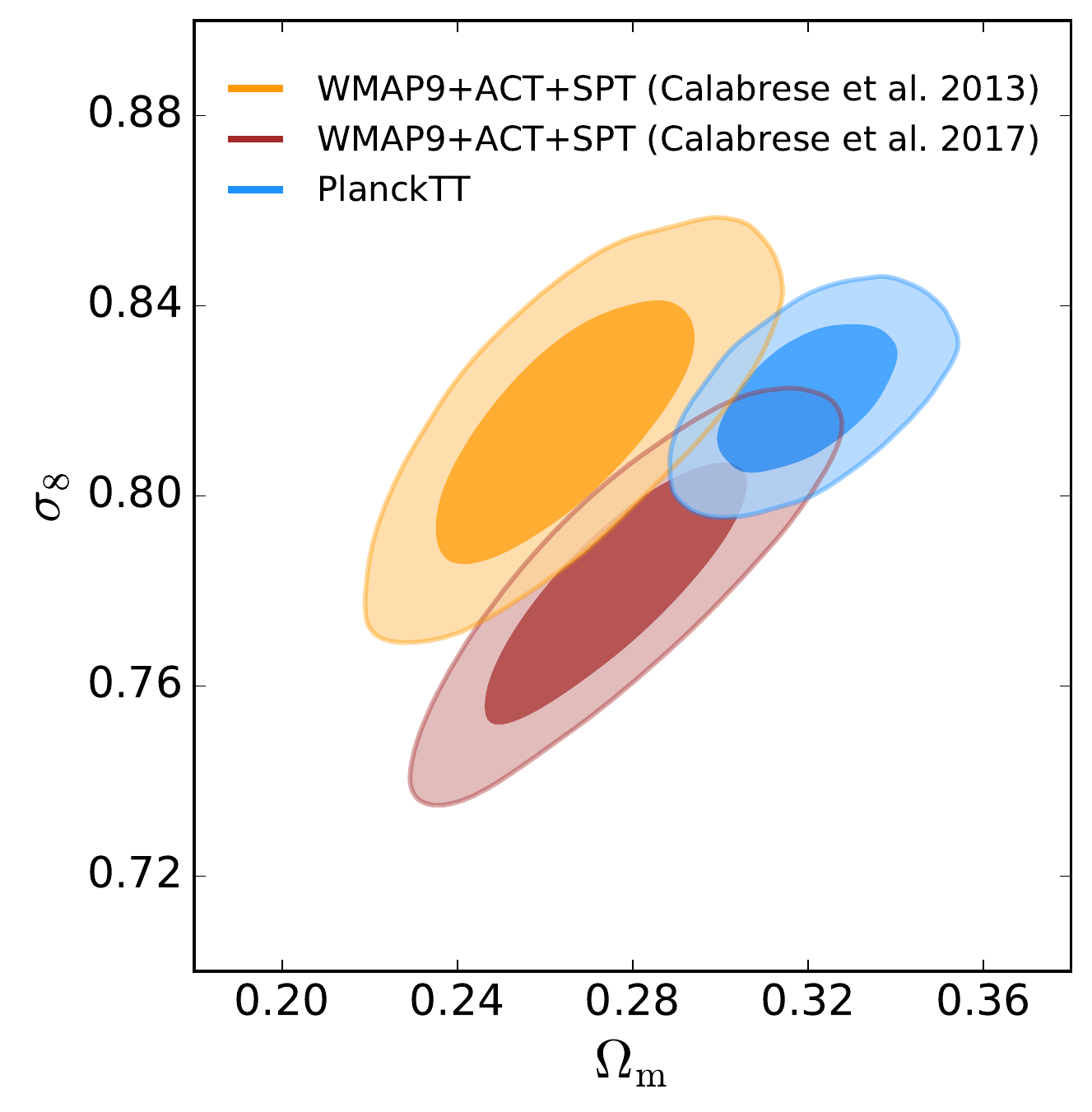}
\caption{Two-dimensional contour regions at 68\% and 95\% confidence for the matter density, $\Omega_m$, and the amplitude of matter fluctuations on 8 $h^{-1} \rm{Mpc}$ scales, $\sigma_8$. A $1.5\sigma$ shift towards lower values of $\sigma_8$ is found in this revised analysis.
\label{fig:2doms8}
}
\end{figure}

As in C13 (to which we refer for details) we make use of the foreground-marginalized ACT+SPT CMB likelihood and combine it with the WMAP public likelihood code. We however do not retain WMAP large-scale polarization and set the flag ``use\_WMAP\_lowl\_pol=F" in the WMAP likelihood options module. We call the likelihoods within the publicly available CosmoMC software~\cite{cosmomc} to estimate the basic six $\Lambda$CDM cosmological parameters: the baryon and cold dark matter densities, $\Omega_b h^2$ and $\Omega_c h^2$, the angular scale of the acoustic horizon at decoupling, $\theta$, the reionization optical depth, $\tau$, the amplitude and the scalar spectral index of primordial adiabatic density perturbations, $A_s$ and $n_s$, both defined at a pivot scale $k_0 = 0.05$ Mpc$^{-1}$. 

We assume a single family of massive neutrinos carrying a total mass of $\Sigma m_\nu=0.06$~eV, and fold in the new \textit{Planck} $\tau$ measurement by imposing a Gaussian prior of $\tau=0.06 \pm 0.01$\footnote{We use a prior because the measurement presented in Ref.~\cite{Plancktau2016} has not been yet accompanied by a likelihood software. This is a conservative choice slightly larger than the \textit{Planck} result of $\tau=0.055\pm0.009$ but leading to the same conclusions.}.
We note that in C13 $\tau$ was measured to be $0.085\pm0.013$ using WMAP polarization at $\ell<23$.

\subsection*{$\Lambda$\rm{CDM}}

The updated constraints on the $\Lambda$CDM basic parameters are reported in Table~\ref{tbl1} and shown in Figure~\ref{fig:lcdm}. A direct comparison between C13 and this revised analysis is reported in column 5 of Table~\ref{tbl1} in terms of shift in cosmological parameters in units of the standard error on that parameter marginalized over the other parameters (basic $\Lambda$CDM posteriors are also compared in Figure~\ref{fig:lcdm}). The largest difference is a $1.9\sigma$ shift of the amplitude parameter which is degenerate with both $\tau$ and the neutrino mass. This shift in $A_s$ will affect most of the derived parameters and in particular the matter density and the amplitude of matter fluctuations. The Hubble constant also moves a non-negligible amount and is $1.7\sigma$ lower than local measurements \cite{Riess2016}. 

Table~\ref{tbl1} also reports the constraints from \textit{Planck} data derived with the same $\tau$ prior replacing the \textit{Planck} low-$\ell$ polarization. 

\subsection*{$\Omega_m$ and $\sigma_8$}

The comparison between CMB and low-redshift constraints is usually reported in terms of the matter density and the amplitude of matter fluctuations on 8 $h^{-1} \rm{Mpc}$ scales, $\Omega_m-\sigma_8$. We want to stress here that the  revised WMAP9+ACT+SPT contours move by more than $1\sigma$ in the $\Omega_m-\sigma_8$ plane compared to C13, as shown in Figure~\ref{fig:2doms8}. This shift receives equal contribution either from having a different, lower $\tau$ or from having non-zero neutrino masses: we find that both parameters move the contours vertically by roughly the same amount. 

These updated contours move along the galaxy weak lensing $\Omega_m-\sigma_8$ degeneracy line and therefore the overall agreement between 2013 CMB data and galaxy shear is unchanged; however, we note that the WMAP9+ACT+SPT contours also overlap at the $1\sigma$ level with the \textit{Planck} CMB contours (when using the same $\tau$ prior) and show better consistency between CMB experiments.


\section{Conclusions}\label{sec:conclusions}

In this short note we have updated the analysis of WMAP, ACT and SPT data presented in Calabrese et al. 2013 (C13) in light of new, commonly used conventions for neutrino masses and a new measurement of the optical depth to reionization from the \textit{Planck} satellite. Unlike in C13, we include as baseline in our cosmological model a single family of massive neutrinos with a total mass of 0.06~eV and impose a prior $\tau=0.06\pm0.01$, replacing the WMAP large-scale polarization information. We show that all basic cosmological parameters shift because of this and highlight the importance of using these revised constraints when comparing different CMB results or when assessing agreement with low-redshift probes.


\acknowledgements
This work was supported by the U.S. National Science Foundation through awards AST-0408698 and AST-0965625 for the ACT project, as well as awards PHY-0855887 and PHY-1214379. Funding was also provided by Princeton University, the University of Pennsylvania, and a Canada Foundation for Innovation (CFI) award to UBC. ACT operates in the Parque Astron\'omico Atacama in northern Chile under the auspices of the Comisi\'on Nacional de Investigaci\'on Cient\'ifica y Tecnol\'ogica de Chile (CONICYT). Computations were performed on the GPC supercomputer at the SciNet HPC Consortium. SciNet is funded by the CFI under the auspices of Compute Canada, the Government of Ontario, the Ontario Research Fund -- Research Excellence; and the University of Toronto. We acknowledge the use of the Legacy Archive for Microwave Background Data Analysis (LAMBDA). Support for LAMBDA is provided by the NASA Office of Space Science. The Dunlap Institute is funded through an endowment established by the David Dunlap family and the University of Toronto. EC is supported by a STFC Rutherford Fellowhisp. 

\bibliography{ref}

\end{document}